\documentclass[epj]{svjour}
\usepackage{graphics}
\begin{document}
\bibliographystyle{prsty}
\title{The $\beta$-decay of $^{22}Al$}
 
\author{N.L.~Achouri\inst{1} \and F.~de~Oliveira Santos\inst{2} \and M.~Lewitowicz\inst{2} \and B.~Blank\inst{3} 
\and J.~\"Ayst\"o\inst{4} \and  G.~Canchel\inst{3} \and S.~Czajkowski\inst{3} \and P.~Dendooven\inst{5} \and A.~Emsallem\inst{6} 
\and J.~Giovinazzo\inst{3} \and N.~Guillet\inst{3} \and A.~Jokinen\inst{4} \and A.M.~Laird\inst{7} \and  C.~Longour\inst{8} \and K.~Per\"aj\"arvi\inst{9} \and N.~Smirnova\inst{10} \and M.~Stanoiu\inst{2,11} \and J-C.~Thomas\inst{3,2} }                
\institute{Laboratoire de Physique Corpusculaire Caen - ISMRA, 6 Bd du Mal Juin, F-14050 Caen Cedex, France. \and 
Grand Acc\'el\'erateur National d'Ions Lourds, Bd Henri Becquerel, B.P. 55027, F-14076 Caen Cedex 5, France. \and
Centre d'Etudes Nucl\'eaires de Bordeaux-Gradignan, Le Haut-Vigneau, F-33175 Gradignan, France. \and
Department of Physics, P.O. Box 35, FIN-40014 University of Jyv$\ddot{a}$skyl$\ddot{a}$, Finland. \and
Kernfysisch Versneller Instituut, Zernikelaan 25, 9747 AA Groningen, Netherlands. \and
Institut de Physique Nucl\'eaire de Lyon, Universit\'e Claude Bernard de Lyon, Villeurbanne, France. \and
Department of Physics, University of York, Heslington, York YO10 5DD, United Kingdom. \and
IReS Strasbourg, 23 rue du Loess, BP 28, F-67037 Strasbourg Cedex, France. \and
Lawrence Berkeley National Laboratory, 1 Cyclotron Road, Berkeley, CA 94720, USA. \and
Vakgroep Subatomaire en Stralingsfysica Universiteit Gent, Belgium. \and
Gesellschaft f$\ddot{u}$r Schwerionenforschung mbH, Planckstr. 1, 64291 Darmstadt, Germany. }

\date{Received: date / Revised version: date}

\abstract{In an experiment performed at the LISE3 facility of GANIL, we studied the decay of $^{22}Al$ produced by the fragmentation of a $^{36}Ar$ primary beam. A $\beta$-decay half-life of $T_{1/2} = 91.1 \pm 0.5~ms$ was measured. The $\beta$-delayed one- and two-proton emission as well as $\beta$-$\alpha$ and $\beta$-delayed $\gamma$ decays were measured and allowed us to establish a partial decay scheme for this nucleus. New levels were determined in the daughter nucleus $^{22}Mg$. The comparison with model calculations strongly favours a spin-parity of $I^\pi = 4^+$ for the ground state of $^{22}Al$. }

\PACS{
     {21.10.-k}{Properties of nuclei; nuclear energy levels}   \and
     {23.90.+w}{Other topics in radioactive decay and in-beam spectroscopy} \and
     {23.50.+z}{Decay by proton emission}
    } 

\maketitle

\section{Introduction}
\label{intro}

Our understanding of the structure of the atomic nucleus is mainly based on studies of stable nuclei or nuclei close
to stability. However, it is now well established that the nuclear structure changes with the addition of
the isospin degree of freedom, i.e. when studying nuclei far away from the valley of stability. The study of these 
exotic nuclei is today a well established tool for a deeper understanding of nuclear structure.

The exotic odd-odd nucleus $^{22}Al$ (Z=13, N=9) is situated at the proton drip line of the nuclear chart and 
is evaluated to be bound by only 20~keV~\cite{audi03}. It was observed for the first time in 1982 by Cable {\it et al.}~\cite{cable82}. The spin of the ground state of $^{22}Al$ is not known. In the mirror nucleus $^{22}F$, 
the spin of the ground state is $I^\pi = 4^+$, but this level is very close to a $3^+$ excited state ($E^*$=71.6~keV) and these 
states could be inverted in $^{22}Al$. As the $\beta$-decay energy of $^{22}Al$ is high ($Q_{\beta^+}~=~17.56$ MeV)~\cite{audi03}, 
different decay channels are open: $\beta$-$\gamma $, $\beta$-p (the one-proton separation energy $S_{p}$ in $^{22}Mg$ is equal to 5501.6~keV), 
$\beta$-2p ($S_{2p}$ = 7931~keV) and $\beta$-$\alpha $ ($S_{\alpha}$ = 8138.7~keV) decays are energetically possible. 

Cable {\it et al.}~\cite{cable82} were the first to study the $\beta$ decay of this nucleus. In their experiment, $^{22}Al$ was produced
by a $^{24}Mg(^3He,p 4n)^{22}Al$ reaction at 110~MeV. The aluminum atoms were transported using a helium-jet technique. In the decay-energy 
spectrum, only two peaks at high energy were clearly seen, the low-energy part being contaminated by the decay of other nuclei.
The measured energies correspond to 8212~$\pm$~16~keV and 8537~$\pm$~22~keV in the center of mass frame. Using energy and intensity considerations, 
these peaks were assigned to the isospin-forbidden proton decay from the lowest T=2 state in $^{22}Mg$ - the isobaric analogue state (IAS) 
of the $^{22}Al$ ground state - to the ground and the first excited states in $^{21}Na$. This IAS is measured to be at an excitation energy of 
13650~$\pm$~15~keV and is fed by a superallowed $\beta^+$ decay. Based on the T=2 isospin multiplet, a $I^{\pi} = 4^+$ was assigned to the 
$^{22}Al$ ground state. A rough half-life value of $70^{+50} _{-35} ms$ was also determined. Due to the technique used, no absolute branching 
ratios could be determined. In another study of this nucleus, the first observation of the exotic $\beta$-2p decay was published by Cable {\it et al.}~\cite{cable83}. Two $\beta$-delayed 2p decay branches were observed. From energy considerations, these two branches were attributed 
to transitions from the IAS in $^{22}Mg$ to the first excited state and the ground state of $^{20}Ne$. In a recent experiment performed by 
Blank {\it et al.}~\cite{blank,czajkowski}, the $\beta$-$\alpha$ decay of this nucleus was observed for the first time using a 
technique based on the implantation of $^{22}Al$ into a silicon detector and into a micro-strip gas counter. 
A half-life of $T_{1/2} = 59 \pm 3~ms$ was measured. A more detailed experimental decay scheme was obtained with absolute branching ratios 
for $\beta$-$\alpha$, $\beta-p$ and $\beta-2p$ emission. However, this experiment suffered from contamination. 
The $^{22}Al$ nuclei were produced by fragmentation of a $^{36}Ar$ primary beam and only about $30\%$ of the nuclei detected were $^{22}Al$ 
nuclei. Shell model calculations were also reported in this publication. From several arguments, the hypothesis that the ground state of 
$^{22}Al$ has a spin-parity of $3^+$ was favoured. Finally, a detailed 1s-0d shell-model calculation was carried out by Brown~\cite{brown} 
for the isospin-forbidden $\beta$-delayed proton emission from the IAS in $^{22}Mg$. A first complete set of calculations for this particle 
decay was presented. 

The $\beta$-decay of this nucleus is quite well known but still suffers from uncertainties and limitations. Due to strong contaminations 
in the previous measurements, we can expect that several transitions are still missing and probably hidden by much stronger transitions from 
contaminants. In addition, the $\gamma$ rays were not measured in any of these experiments. The aim of our experiment was to improve the 
$\beta$-decay measurement by means of different advantages: a better purity of the $^{22}Al$ secondary beam, higher statistics, greatly 
improved energy resolution in the measurement of the charged particles emitted, a $\gamma$-ray measurement and $\gamma$-particle coincidences 
to give a more certain assignment for many transitions. 

A comparison of the experimental results with detailed shell model calculations 
is also proposed in the present paper. For the first time, a complete set of calculations for all particle decay channels from 
relevant levels in $^{22}Mg$ is done. This will allow us to investigate the Gamow-Teller quenching in $\beta$-decay, to test 
the isobaric multiplet mass equation (IMME), and to probe the isospin symmetry in $\beta$-transitions of mirror decays. Another 
challenge is to determine the spin of the $^{22}Al$ ground state.

\section{The experimental setup}
\label{sec:2}

The study of the $^{22}Al$ $\beta$ decay was performed at GANIL with the LISE3 zero-degree achromatic recoil spectrometer~\cite{lise,lise3}. 
It was part of a systematic $\beta$-decay study of neutron deficient nuclei: $^{27}S$~\cite{canchel}, $^{25}Si$ and $^{26}P$~\cite{thomas} 
where the same experimental setup was used. The nucleus $^{22}Al$ was produced by the fragmentation of a 
95~A~MeV $^{36}Ar$ primary beam on a 528 mg/cm$^2$ carbon target located in the 
SISSI device. The $^{22}Al$ secondary beam had an energy of 48~A~MeV. 
Using a beryllium degrader at the intermediate LISE focal plane and the LISE3 Wien filter, this isotope was produced 
with an intensity of 40 pps and a purity of $93\%$. The main contaminants were $^{21}Mg$ (1.5\%), $^{20}Na$ 
(3.5\%) and $^{19}Ne$ (2\%). A total number of $2.5\times10^6$ $^{22}Al$ was recorded in this experiment. 
Other settings of the spectrometer were meant to select and measure the $^{21}Mg$ and $^{24}Al$ $\beta$ decays 
with the same detection system for energy and efficiency calibration. 

\begin{figure}
\resizebox{0.5\textwidth}{!}{%
  \includegraphics{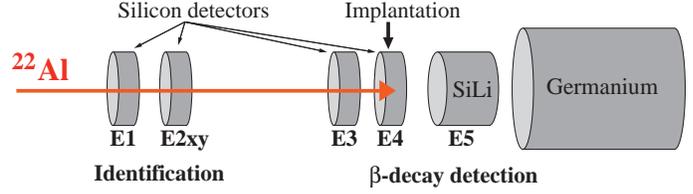}
}
\caption{\it Experimental setup composed of two sets of silicon detectors and a germanium clover in close geometry. The first 
two silicon detectors served for the identification of $^{22}Al$ nuclei (E1 and E2xy), whereas the other detectors were used for 
the measurement of the charged particles and the $\gamma$-rays emitted in $\beta$-decay (E3, E4, E5 and the germanium detector). 
The $^{22}Al$ ions were implanted in the silicon detector E4.}
\label{fig:dispositif}
\end{figure}

The experimental setup shown in figure~\ref{fig:dispositif} consisted of two sets of detectors. The identification of the incident 
ions was performed with two $300~\mu m$ silicon detectors (E1 and E2xy) generating an energy-loss signal and a 
multi-channel plate (MCP) detector situated at the first LISE focal plane which, together with the E1 detector, 
produced a time-of-flight signal. 
The $\beta$-decay detection system was composed of two $500~\mu m$ silicon 
detectors (E3, E4), a $6~mm$ Si(Li) detector (E5) and an EXOGAM Germanium clover detector in close geometry. 
The fragments were implanted at the downstream edge of the fourth silicon detector E4, which served also to measure the 
charged-particle decay-energy spectrum. The energy calibration of the E4 detector was performed using the known $\beta$-proton 
transitions of $^{21}Mg$~\cite{sextro}. The E3 and E5 detectors were used as $\beta$-particle detectors.

\section{Monte Carlo simulation}
\label{sec:3}

A Monte Carlo simulation of the experiment was performed using the GEANT code~\cite{geant}. 
This simulation has produced four main conclusions about:

\begin{itemize}

\item The proton detection efficiency: \\
High-energy protons can escape from the E4 implantation detector. The simulation allowed us to extract the proton detection 
efficiency of the E4 detector as a function of proton energy assuming an isotropic proton emission distribution. For energies 
from $2$ MeV to $8$ MeV, the efficiency decreases from 100\% to 52\% for an implantation depth of $420 \pm 10~\mu m$. The 
relative uncertainty of the proton detection efficiency depends on the uncertainty of the implantation depth in the silicon 
detector and on the uncertainty of the detector thickness. After a simulation using the extreme values of the implantation 
profile, the relative uncertainty of the proton detection efficiency was found to be less than 6\%. \\

\item The proton energy resolution: \\
This simulation and previous experiments have shown that a condition on the last silicon detector (E5 \textgreater 0) leads 
to a better energy resolution in the decay-energy spectrum. In figure~\ref{fig:pro_bet}, we have simulated the $\beta$ decay 
of $^{21}Mg$ to one state at 4.468~MeV in $^{21}Na$ followed by the emission of a 2.036~MeV proton. In the E4 spectrum, we can see that the $\beta$ particle associated with the transition has a strong effect. The proton peak is shifted by 36 keV when the positrons escape on the downstream side of E4, and by 158 keV when they escape in the upstream direction. 
These results do not depend on the $Q_{\beta}$ value in the 4-14~MeV range of interest, however, they depend strongly on the implantation depth in the E4 detector. If we use a condition on E5, we can mainly select the $\beta$ particles which escape on the downstream side. Such a condition selects the first peak and gets rid of the second one as shown by the dotted line in figure~\ref{fig:pro_bet}. This technique leads to a much better energy resolution. \\

\item The branching ratio determination: \\
In order to determine the proton branching ratios, we have to determine the area of each proton peak without any condition. However, 
it is difficult to fit the area of the proton peaks with a complex function that takes into account the exact shape of the 
double-bump distribution (see figure~\ref{fig:pro_bet}). Moreover, simulations have shown that the exact shape of the peak depends on the $\beta$ particle energy, which is most often not known. 

\begin{figure}
\resizebox{0.4\textwidth}{!}{%
  \includegraphics{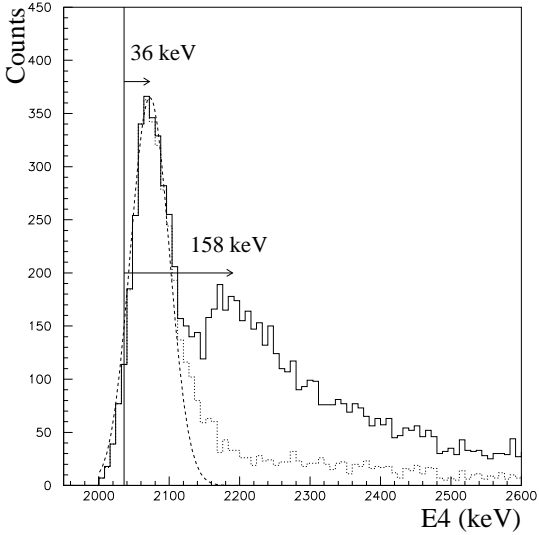}
}
\caption{\it Simulation of a $^{21}Mg$ $\beta$-delayed proton decay in the E4 detector. A decay energy of 2.036~MeV is used 
and is indicated in this figure by the vertical line. The full line represents the spectrum without any condition and shows 
two peaks shifted from the initial position by 36 keV for the first peak due to the energy loss of the $\beta$ particles that 
escape on the downstream side of E4 and by 158 keV when they escape in the other direction. The dotted line represents the proton spectrum conditioned by a $\beta$ detection in E5 (E5 \textgreater 0). The dashed line is the Gaussian fit of this conditioned spectrum peak.}
\label{fig:pro_bet}
\end{figure}

So, to extract the peak area we have used the method described here. First of all, we performed a Gaussian fit of the low energy 
part of the simulated proton spectrum, conditioned by the E5 detector (E5 \textgreater 0) as shown by the dashed line in 
fig.~\ref{fig:pro_bet}. This fit allowed us to calculate the corresponding area. By normalising this area with the total 
peak area without any condition (the total number of simulated particles for this proton energy), we obtain the ratio of 
events selected by such a procedure. We have introduced two ratios:

\begin{enumerate} 

\item $R_1 = \frac{N_{p (E5 \textgreater 0)}}{N_{p total}}$ \\
 
This ratio is determined by dividing the total number of counts in the conditioned proton spectrum (E5 \textgreater 0) by the 
total number of counts in the proton spectrum without any condition. However, the charged-particle spectrum contains a $\beta$ 
contribution from $^{22}Al$ $\beta$-$\gamma$ transitions. 
To get the total number of protons, one has to subtract this contribution from the spectrum. In fact, simulations have shown 
that the $\beta$ contribution can be fitted by the sum of two exponential functions and subtracted from the unconditioned proton 
spectrum. \\

\item $R_2 =\frac{N_{p (E5 \textgreater 0) Gaus}}{N_{p (E5 \textgreater 0)}}$ \\ 
$R_2$ is the ratio between the Gaussian area of the proton peak and the total proton peak area in the conditioned spectrum. 
This takes into account that the Gaussian does not perfectly describe the peak in the conditioned spectrum. 

\end{enumerate}

In practice, the low-energy part of the proton peak in the conditioned spectrum (E5 \textgreater 0) is fitted by a Gaussian. 
The area obtained is divided by $R_2$. This yields the total area of the proton peak in the conditioned spectrum. This 
value is divided by $R_1$ to obtain the area of the total proton peak (without condition). Finally, the branching ratio 
is determined by dividing this area by the total number of $^{22}Al$ implanted. \\
The simulations showed that the ratios $R_1$ and $R_2$ depend weakly on the $\beta$-particle energy (a maximum of 10\% 
relative error). However, since they depend on the geometry, we need to determine these ratios from experimental data at 
least for one proton peak (for one $\beta$-particle energy) and use these ratios for the determination of the other proton 
peak areas. \\

\item The $\beta$-trigger efficiency:  \\
In the present experiment, the acquisition system was triggered by charged particles detected in E3, E4 or E5. If the 
$\beta$ particle 
is followed by proton emission, the energy loss is always sufficient to trigger the acquisition. Otherwise, in the case 
of $\beta$-$\gamma$ 
transitions, the energy lost by $\beta$ particles in E3 or E4 is not sufficient to trigger the acquisition. In this case, the 
acquisition may be triggered by $\beta$ particles detected in the E5 detector.  This efficiency must be determined to calculate 
the branching ratio for the $\beta$-$\gamma$ transitions.

To study the behaviour of the $\beta$-trigger efficiency, a simulation was performed. It has shown that this efficiency depends strongly 
on the distance between the detectors E4 and E5 (i.e. on the solid angle). However, for a fixed geometry, the $\beta$-trigger efficiency 
depends linearly with the $\beta$ energy. For the 4-14~MeV range of interest, the simulation shows that the $\beta$-trigger efficiency 
varies from 36\% to 44\%. 

\end{itemize}

\section{Experimental results}
\label{sec:4}

In the following, we present our experimental results: half-life, $\beta$-delayed $\gamma$ decays, $\beta$-delayed charged-particle 
emissions as well as coincidences between charged particles and $\gamma$ rays and $\gamma$-$\gamma$ coincidences. 
These results allow us to determine a detailed experimental decay scheme and to compare it with previous results. \\

\begin{enumerate}

\item $^{22}Al$ half-life \\

The $^{22}Al$ half-life was measured in beam-on/beam-off mode (120~ms of implantation, 300~ms beam-off). The inset in figure \ref{fig:temps_vie} 
shows the decay-time spectrum in the beam-off period. This spectrum was fitted with an exponential function for the $^{22}Al$ decay and a constant for the background. The decay of the daughter $^{22}Mg$ is the main contributor to this background ($T_{1/2}=3.857~s$). The half-life obtained with this method is $T_{1/2}=~91.9 \pm 1.4~ms$.

\begin{figure}
\resizebox{0.5\textwidth}{!}{%
  \includegraphics{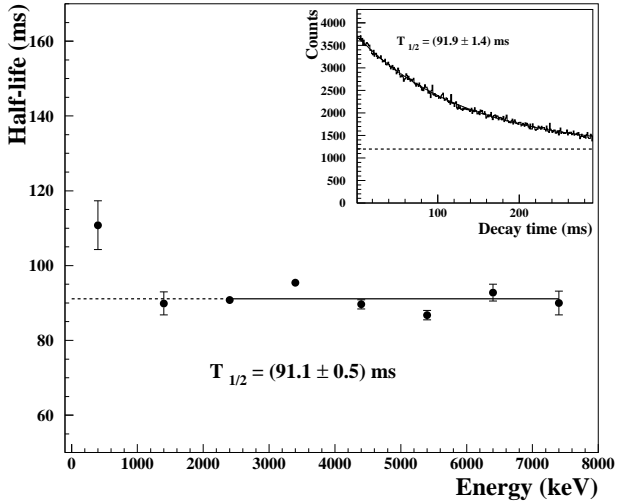}
}
\caption{\it The inset shows the decay-time spectrum in the beam-off period fitted by an exponential function added to a constant for the background. In the main figure, the half-lifes obtained from the fit of different decay-time spectra in coincidence with protons are drawn as a function of proton energies. The half-life given in the main figure is the recommended value (see text). }
\label{fig:temps_vie}
\end{figure}

A second analysis was performed by generating, from the same data set, the decay-time spectra in coincidence with the different proton groups. Each point in the figure~\ref{fig:temps_vie} represents in y axis the half-life obtained from the fit of a decay-time spectrum in coincidence with a range of proton energies in x axis. The error-weighted mean value for the half-life from this analysis is $T_{1/2}=~91.1 \pm 0.5~ms$. The two points at low energy were not used to avoid the contribution from the contaminants. In fact, the main contaminant $^{22}Mg$ decays only by $\beta-\gamma$ and the $\beta$ particles associated contributes to the proton spectrum at low energies. However, this value is more accurate than the previous one which contains a higher contamination. 

A third analysis with a continuous implantation mode has produced a value of $T_{1/2} =~87.3 \pm 1.1~ms$.

Since the second method is less subject to errors than the other methods, we recommend the value of $T_{1/2}= 91.1 \pm 0.5~ms$ as the half-life of $^{22}Al$. \\

\item Beta-delayed charged particle decay \\

\begin{figure*}
\begin{center}
\resizebox{0.75\textwidth}{!}{%
  \includegraphics{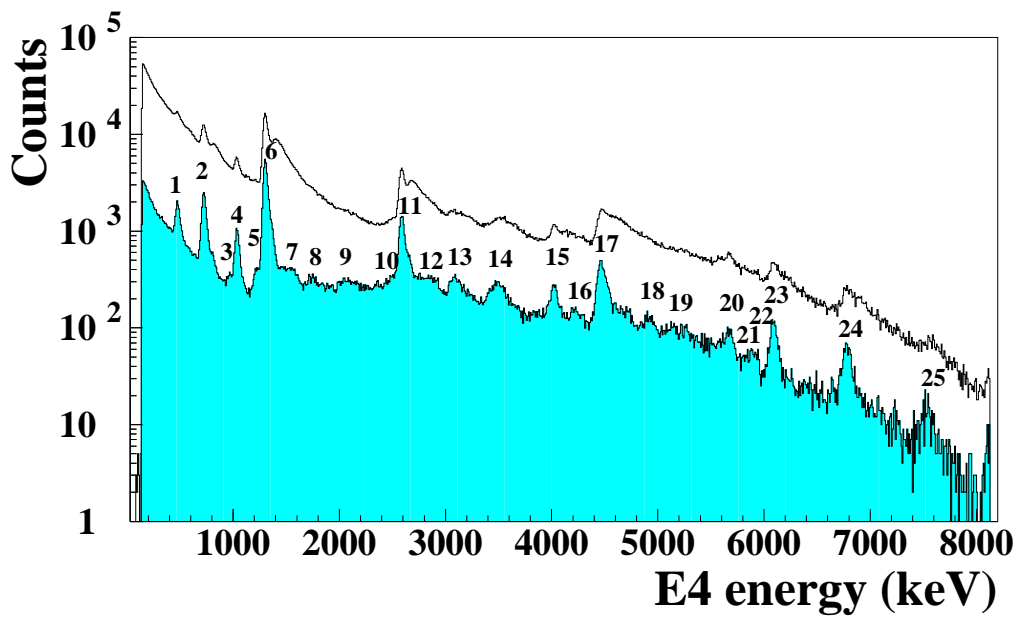}
}
\caption{\it Energy spectrum of the $\beta$-delayed charged particles detected in the E4 silicon detector. The upper spectrum is the unconditioned spectrum, whereas the shaded spectrum is generated with the condition that the E5 detector is triggered as well.}
\label{fig:pro22al}
\end{center}
\end{figure*}

After implantation of $^{22}$Al in the E4 detector, an energy spectrum of $\beta$-delayed charged particles was obtained as shown in 
figure \ref{fig:pro22al}. 
The figure shows the spectrum without any condition and, shaded, the spectrum in coincidence with $\beta$ particles detected in the 
last detector. 
As discussed in the simulation section (section~\ref{sec:3}), this condition provides a better energy resolution of the proton peaks. 
Up to 25 peaks are visible. The 
energy measured is the total decay energy which was corrected for the pulse-height defect~\cite{ratkowski}.

The branching ratio for a transition is defined as the total area of the proton peak in the unconditioned spectrum divided by the 
total number of $^{22}Al$ implanted. The areas of the proton peaks were determined using the method described in the simulation 
section (see section~\ref{sec:3}). \\
First, we determined the two ratios $R_1$ and $R_2$. The ratio $R_1=32.7\%$ was determined by dividing the total number of counts 
in the conditioned proton spectrum (E5 \textgreater 0) by the total number of counts in the unconditioned proton spectrum after 
subtraction of the $\beta$ contribution. \\
The ratio $R_2=53.5\%$ is determined from a fit of a relatively isolated proton peak in the conditioned proton spectrum 
(E5 \textgreater 0). We have chosen the peak labelled 11 at E=2.583~MeV. The low energy part of the peak was fitted by a 
Gaussian while the total peak was fitted by a Gaussian added to an exponential function for the high-energy tail.

Each proton peak in the conditioned spectrum was fitted by a Gaussian, the area obtained was divided by $R_1$ and $R_2$ and by the number of $^{22}Al$ implanted. The branching ratios obtained were corrected for the proton detection efficiency. 

The proton peak energies and their branching ratios are given in table \ref{tab:pro22al} and compared with previous measurements of 
Blank {\it et al.}~\cite{blank}. We observe an overall agreement but many new or better defined peaks above the proton threshold in $^{22}Mg$ were 
measured. \\
The branching ratios of the peaks $15^*$,$17^{**}$ and $23^{**}$ have not been corrected for proton detection efficiency as the other 
peaks because, as it will be shown later, peak $17^{**}$ is assigned to a $\beta$-2p transition to the $^{20}Ne$ first excited state 
and peak $15^*$ corresponds to a $\beta$-$\alpha$ decay to the $^{18}Ne$ first excited state. In these two cases, the $\beta$-2p 
and $\beta$-$\alpha$ peaks are superimposed on proton transitions. The only way to distinguish between these transitions and the proton 
emission, to get the branching ratios, is to use the coincidences with $\gamma$-rays. \\
The peak $23^{**}$ with $E=6.085 \pm 0.008$ MeV is compatible with the $\beta$-2p decay measured by Cable {\it et al.}~\cite{cable83} 
to the $^{20}Ne$ ground state. For the $\beta$-2p transitions, the proton detection efficiency was taken at half of the transition energy. \\

\begin{table*}
\begin{center}
\begin{tabular}{|c|c|c|c|c|}
\hline
 Peak & \multicolumn{2}{|c|}{This work}& \multicolumn{2}{|c|}{Blank {\it et al.} \cite{blank}} \\
\cline{2-5} 
    &  Energy($MeV$) & $Br$($\%$)     &  Energy($MeV$) & $Br$($\%$) \\
\hline
 1 & 0.475$\pm$0.008 & 4.73$\pm$ 0.63 &  0.45$\pm$0.04  & 6.4$\pm$1.2 \\
\hline
 2 & 0.721$\pm$0.008 & 7.39$\pm$ 1.01 &  0.72$\pm$0.04  & 6.8$\pm$1.2 \\
\hline
 3 & 0.975$\pm$0.008 & 0.25$\pm$ 0.05 &                 &  \\
\hline 
 4 & 1.033$\pm$0.008 & 3.00 $\pm$ 0.34 &  1.04$\pm$0.04  & 3.9$\pm$1.2 \\
\hline
 5 & 1.223$\pm$0.008 & 0.75$\pm$ 0.10 &                 &    \\
\hline
 6 & 1.299$\pm$0.008 & 18.51$\pm$1.74 & 1.32$\pm$0.04   & 18.0$\pm$1.0\\
\hline
 7 & 1.551$\pm$0.010 & 0.81$\pm$ 0.16 &                 &             \\
\hline
 8 & 1.753$\pm$0.008 & 0.45$\pm$ 0.08 &                 &   \\
\hline
   &                 &                &  1.95$\pm$0.06  & 3.2$\pm$1.0 \\ 
\hline
 9 & 2.072$\pm$0.008 & 0.48$\pm$ 0.07 &                 &   \\
\hline
 10 & 2.503$\pm$0.010 & 0.64$\pm$0.13 &                 &   \\
\hline
 11 & 2.583$\pm$0.008 & 4.89$\pm$0.24 &                 &   \\
\hline
 12 & 2.838$\pm$0.008 & 2.11$\pm$0.09  &                &  \\
\hline
 13 & 3.088$\pm$0.008 & 1.89$\pm$0.07  &                &  \\
\hline 
 14 & 3.484$\pm$0.008 & 2.18$\pm$0.15  &                &  \\
\hline
 15 & 4.017$\pm$0.008 & 1.04$\pm$0.33  &                &  \\
$15^*$& 4.017$\pm$0.008 & 0.038$\pm$0.017 & 3.997$\pm$0.049& 0.31$\pm$0.09 \\
\hline
   &                 &                & 4.10 $\pm$ 0.06 & 2.1$\pm$1.0\\ 
\hline
 16& 4.224$\pm$0.009 & 0.84$\pm$ 0.11      &            &    \\
\hline
 17    & 4.464$\pm$0.008& 2.52$\pm$0.14 &   &    \\
$17^{**}$& 4.464$\pm$0.008& 0.69$\pm$0.08 & 4.48 $\pm$0.25& 0.9 $\pm$ 0.5   \\
\hline
 18& 4.912$\pm$0.010& 0.27$\pm$0.32   & 4.92$\pm$0.07   & 2.4 $\pm$ 1.0 \\       \hline 
 19& 5.177$\pm$0.013& 0.29$\pm$0.11   &             &  \\
\hline 
 20& 5.667$\pm$0.008& 0.35$\pm$0.11   &             &  \\
\hline
 21& 5.808$\pm$0.049 & 0.18$\pm$0.55  &                 &  \\
\hline
 22& 5.909$\pm$0.056 & 0.21$\pm$0.62  &                 &  \\
\hline
 23$^{**}$& 6.085$\pm$0.008& 0.41$\pm$0.07 &            &  \\
\hline
 24& 6.774$\pm$0.008 & 0.41$\pm$0.12  &  6.62$\pm$0.10  & 0.7$\pm$0.3  \\
\hline
   &                 &                & 6.93$\pm$0.10   & 0.1$\pm$0.05 \\
\hline
 25& 7.517$\pm$0.011 & 0.33$\pm$0.07  &                 &              \\
\hline
   &                 &                & 8.29$\pm$0.10   & 0.2$\pm$0.05 \\
\hline
\end{tabular}
\end{center}
\caption{\it Total decay energies of $\beta$-delayed protons and their branching ratios corrected for the proton detection efficiency except for the 
transitions labelled by * and by **. The transition labelled by * is a $\beta$-$\alpha$ transition superimposed on a proton transition. The branching ratio of the $\alpha$ transition was derived from the intensity of a $\gamma$-ray observed in coincidence (see text). The two transitions labelled by ** are $\beta$-2p transitions (see text). The present results are compared to a previous measurement by Blank {\it et al.}~\cite{blank}.}     
\label{tab:pro22al}
\end{table*}

\item $\gamma$-ray spectrum \\

\begin{figure}
\resizebox{0.5\textwidth}{!}{%
  \includegraphics{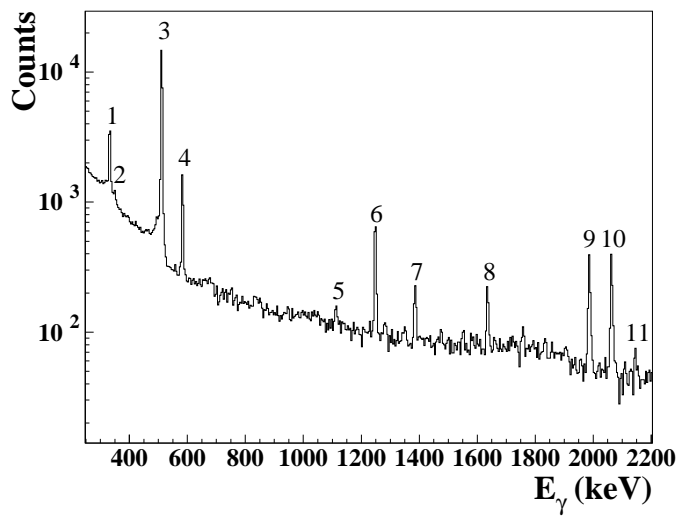}
}
\caption{\it $\gamma$-ray spectrum detected in the germanium clover detector calibrated in energy. This spectrum is the sum of the four spectra measured in each crystal. The origin of the different $\gamma$ rays is discussed in the text.}
\label{fig:gamma}
\end{figure}

Figure \ref{fig:gamma} shows eleven $\gamma$ rays detected in the germanium clover detector during the $^{22}Al$ measurement. Only six of them (labelled 2, 4, 6, 9, 10 and 11) are in coincidence with a low-energy event in the E4 detector (E4 \textless~400~keV). This energy range corresponds mainly to $\beta$ particles emitted without a coincident proton. The other $\gamma$ rays 1, 5, 7 and 8 will be disscussed below. \\
No $\gamma$-$\gamma$ coincidences could be established using the diagonally opposite detectors of the clover. This lack of coincidences does not mean that there are no $\gamma$ cascades, but it is rather due to the relatively low efficiency of each individual segment of the germanium detector and to low statistics. \\

\begin{itemize}

\item $\beta$-$\gamma$ decay \\

In figure~\ref{fig:gamma}, the $\gamma$ ray 2 at $351 \pm 2$ keV comes from the $\beta$-$\gamma$ decay of $^{21}Na$ \cite{firestone04}. The $\gamma$ ray 4 at $583.7 \pm 1.6$ keV corresponds to the $\beta$-$\gamma$ decay of $^{22}Mg$ \cite{firestone05}. 
The other $\gamma$-rays 6, 9, 10, 11 at $1248.5 \pm 2.0$ keV, $1985.6 \pm 1.3$ keV, $2062.3 \pm 1.5$ keV and $2145 \pm 5$ keV are attributed to the $\beta$-$\gamma$ decay of $^{22}Al$. In fact, four $\gamma$-ray transitions at $1246.98 \pm 0.03$ keV, $1984.8 \pm 0.1$ keV, $2061.09 \pm 0.05$ keV and $2143.5 \pm 0.6 $ keV were previously established in $^{22}Mg$ (see~\cite{firestone96},~\cite{firestone05}). The $\gamma$ rays 6 and 10 should correspond to these known transitions. However, according to the reference~\cite{firestone96}, a known $\gamma$ ray at 1984~keV should be accompanied by another $\gamma$ ray at 891~keV with a relative branching ratio of 67\%. From 657 counts observed in the $\gamma$ peak at 1985.6 keV we would expect 837 counts at 891 keV. This is not the case. 
We deduce therefore that the $\gamma$ ray 9 measured at 1985.6 keV is compatible  with the new adopted $\gamma$ transition in reference~\cite{firestone05} with 100\% relative branching ratio. From our measurement, this $\gamma$ ray is most likely emitted from a level situated at $5.294 \pm 0.003$~MeV in $^{22}Mg$. This assumption is in agreement with the adopted value of $5.29311 \pm 0.00015$~MeV and supported by theoretical calculations (see section~\ref{sec:6} below). The weak $\gamma$-ray labelled 11 in fig.~\ref{fig:gamma} is situated at $2145 \pm 5$ keV. 
It may correspond to a known $\gamma$-ray in $^{22}Mg$ measured at $2143.5$ keV emitted from a level situated at $5.4524 \pm 0.0004$~MeV~\cite{firestone05}. 
This level also emits another $\gamma$-ray at $4205.4$ keV. Since our energy range is limited to $3800$ keV, this fact can not be checked. 
We assume that the $\gamma$-ray 11 comes from this known level. The absolute intensities of the $\gamma$ transitions are given in table~\ref{tab:intens_bet_gam} corrected for $\gamma$ efficiency and $\beta$-trigger efficiency.   
The $\beta$-trigger efficiency has been determined from the well known $\beta$-$\gamma$ transitions of $^{24}Al$, also produced in 
this experiment for detector calibration. Two $^{24}Al$ $\gamma$ rays at 1077~keV and 1368~keV \cite{firestone96} have been used and lead to a value 
of $39 \pm 3$\% for the $\beta$-trigger efficiency. \\

\begin{table}
\begin{center}
\begin{tabular}{c|c|c}
\hline
Energy (MeV)            &  Intensity (\%)   & Parent nucleus       \\
\noalign{\smallskip}\hline\noalign{\smallskip}                                    $583.7 \pm 1.6$        &   $47.4 \pm 9.3$  & $^{22}$Mg     \\
\hline                  
 $1248.5 \pm 2.0$       &   $38.2 \pm 6.9$  & $^{22}$Al     \\
\hline                  
 $1985.6 \pm 1.3$       &   $31.1 \pm 5.4$  & $^{22}$Al     \\
\hline                  
 $2062.3 \pm 1.5$       &   $34.1 \pm 5.8$  & $^{22}$Al     \\
\hline                                     
 $2145 \pm 5$           &   $1.7 \pm 0.7$   & $^{22}$Al     \\
\hline                                     
\end{tabular}
\end{center}
\caption{\it Measured energies and absolute intensities of $\gamma$-rays from the $\beta$-$\gamma$ decay of $^{22}Al$. The intensities 
are corrected for the detection and the trigger efficiencies. The first $\gamma$ ray comes from the decay of $^{22}Mg$ and corresponds 
to the total $\beta$-$\gamma$ decay probability of $^{22}Al$.}   
\label{tab:intens_bet_gam}
\end{table}


\item $\gamma$-particles coincidences \\

Four $\gamma$-rays noted 1, 5, 7 and 8 in figure \ref{fig:gamma} were measured at $332 \pm 1.2$ keV, $1112.9 \pm 2.4$ keV, $1385.5 \pm 1.3$ keV and $1633.8 \pm 2.2$ keV. The three first rays are known as $\gamma$ transitions in $^{21}Na$ \cite{firestone04}, which suggest a $\beta$-proton-$\gamma$ decay from $^{22}Al$ to $^{21}Na$. The E4 spectrum conditioned by these 3 $\gamma$ rays 
shows coincidences. 
We have observed that the $\gamma$ ray at 332~keV is in coincidence with proton lines 1, 4, 6 and 11 (figure~\ref{fig:pro22al}). The 
coincidence spectrum with the $\gamma$ ray at 1112.9~keV is not conclusive. Moreover, the $\gamma$ ray at 1385.5~keV is in coincidence 
with the proton peak 20. 

The charged-particle peak $17^{**}$ with $E = 4.464 \pm 0.008$ MeV is compatible with the $\beta$-2p decay energy 
measured by Cable {\it et al.}~\cite{cable83} and Blank {\it et al.}~\cite{blank} at $E = 4.48 \pm 0.06$ MeV. The $\gamma$-ray spectrum 
observed in coincidence with this peak is shown in figure~\ref{fig:coinc_2p}. It clearly shows a coincidence with a $\gamma$ ray at 1633.8~keV, 
corresponding to the transition from the first excited state to the ground state in $^{20}Ne$. This is a firm confirmation of the $\beta$-2p 
decay to the $^{20}Ne$ first excited state. The branching ratio of this $\beta$-2p decay is calculated from the number of counts in the 2p 
peak in the charged-particle spectrum conditioned by the $\gamma$ ray at 1633.8~keV. After correction for the $\gamma$ and proton detection 
efficiencies and normalisation to the total number of $^{22}Al$ implanted, we find a branching ratio of $0.69 \pm 0.08 \%$. 

\begin{figure}
\resizebox{0.4\textwidth}{!}{%
  \includegraphics{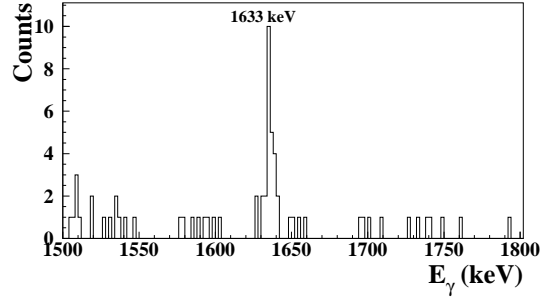}
}
\caption{\it Gamma energy spectrum in coincidence with the charged-particle peak at 4.464 MeV in the E4 detector. The coincidence 
of this charged-particle peak with the 1633.8~keV $\gamma$ ray from $^{20}Ne$ demonstrates that the peak $17^{**}$ (at 4.464 MeV) is a $\beta$-2p 
decay to the first excited state in $^{20}Ne$.}
\label{fig:coinc_2p}
\end{figure}
\begin{figure}
\resizebox{0.4\textwidth}{!}{%
  \includegraphics{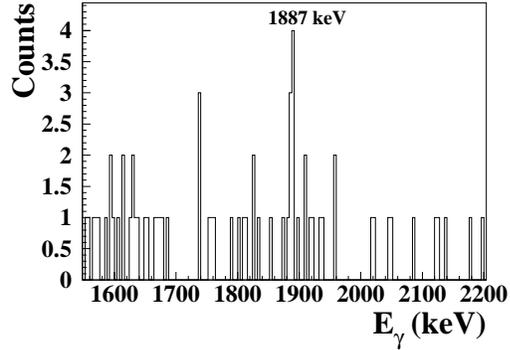}
}
\caption{\it Gamma energy spectrum in coincidence with the charged-particle peak $15^{*}$ at 4.017 MeV in the E4 detector which confirms the 
$\beta$-$\alpha$ transition to the first excited state of $^{18}Ne$.}
\label{fig:coinc_alpha}
\end{figure}

The energy of the charged-particle peak $15^*$ ($E=4.017 \pm 0.008$ MeV) is compatible with the $\beta$-$\alpha$ transition measured 
by Blank {\it et al.}~\cite{blank} at $E = 4.01 \pm 0.05$ MeV. The $\gamma$ spectrum conditioned by this peak shows a clear $\gamma$ ray 
at 1887 keV (see figure~\ref{fig:coinc_alpha}). This $\gamma$-ray energy corresponds to the $\gamma$ transition from the first excited 
state to the $^{18}Ne$ ground state. Again, this observation is the first firm confirmation of this $\beta$-$\alpha$ transition. 
The branching ratio of the $\beta$-$\alpha$ decay to the first excited state of $^{18}Ne$ is $0.038 \pm 0.017~\%$. 

\end{itemize}

\end{enumerate}

\section{Shell model calculations}
\label{sec:5}

Shell model calculations were performed to investigate the $\beta$-decay of $^{22}Al$. 
One aim of these calculations was to determine the branching ratios to the different states in $^{22}Mg$. Different calculations 
already exist for this isotope~\cite{blank,brown}. We have performed new calculations with an isospin non-conserving 
interaction. No significant difference has been observed compared to an isospin-conserving calculation.

These calculations were also performed for the following reason. To construct an experimental decay scheme, we have to interpret the 
different experimental decay peaks in terms of their absolute energy position as states in $^{22}Mg$. This procedure is subject to 
uncertainties, because we deal with a large number of peaks and a huge number of states and transitions. It is obvious that errors 
in the interpretation can result in an apparent change of the Gamow-Teller strength. To avoid this problem, we have determined the different branching ratios for the emission of $\gamma$ rays, protons and sequential two protons from $^{22}Mg$ states. Brown~\cite{brown} has already done this type of calculations in the case of $^{22}Al$ for the particle emission 
from one unique state, the IAS in $^{22}Mg$. For the first time, we have calculated these branching ratios for all interesting states 
in $^{22}Mg$, from the ground state to levels up to a $Q_{\beta^+}$ = 17.56 MeV. Thus, we are able to compare directly the experimental 
spectrum with the predicted one using a comparison between the branching ratios measured and those calculated 
(see figure~\ref{fig:sum_br}).

We performed $sd$ shell-model calculations using the OXBASH code~\cite{oxbash}. Two different interactions were used, 
the USD interaction~\cite{wild} and the isospin non-conserving Ormand-Brown OB interaction~\cite{ormand}. Only the OB 
interaction allowed us to calculate the particle decay from the IAS in $^{22}Mg$, because it is isospin forbidden. Since 
the ground state spin of $^{22}Al$ is not known, we performed two calculations, one for a spin-parity of $3^+$ and one for 
a spin-parity of $4^+$. For the $\beta$ transitions, we used effective single-particle matrix elements -an overall quenching factor of $0.76 \pm 0.03$
is included- for the Gamow-Teller operator~\cite{brown85}. 

The phase-space factor $f$ for $\beta ^+$-decay was calculated by means of a program available at~\cite{nndc}. For the particle transitions, 
it is necessary to calculate the spectroscopic factors $\theta ^2$ for all possible transitions from states in $^{22}Mg$ to states in the daughter 
nuclei. The partial widths were estimated in the standard way from the expression $\Gamma = 2 \theta ^2 \gamma ^2 P(l,Q)$ where $Q$ is the 
particle-decay energy, $\gamma ^2$ is the Wigner single-particle width, $P(l,Q)$ is the penetrability, and $l$ is the angular momentum of the 
transition. The penetrabilities are calculated in a Woods-Saxon well using the correct number of nodes of the wave functions.

The calculated decay schemes are similar to those presented in~\cite{blank}. The calculated $^{22}Al$ half-life is 80~ms in the $3^+$ case 
and 85~ms in the $4^+$ case. It is similar to a previous study~\cite{brown}, where half-lives of $T_{1/2} = 78~ms$ for a $3^+$ state 
and of $T_{1/2} = 85~ms$ for a $4^+$ state were found. The difference between $3^+$ and $4^+$ calculations is not enough significant to prefer one or the other hypothesis. 

However, the differences in the calculated branching ratios are more significant. For a $4^+$ ground state of $^{22}Al$, the branching 
ratio of 
$\beta$-$\gamma$ decay is 39.4\%, whereas the same branching ratio is 48.9\% for a $3^+$ ground state. Similarly, a $\beta$-p branching 
ratio of 53\% is obtained in the $4^+$ case, while this number drops to 39.5\% in the $3^+$ case.

$\beta$-2p emission has also been investigated as a sequential emission, a first proton is emitted from $^{22}Mg$, followed 
by a second proton emitted from an excited state in $^{21}Na$. If we consider each transition individually, weak branching ratios 
are obtained. Only one transition could be measured by our setup and corresponds to a branching ratio of 0.84\% for a total proton 
energy of $4.215$ MeV for a $4^+$ ground state of $^{22}Al$. However, our setup can not distinguish between transitions to different 
intermediate states in $^{21}Na$. The measured energy is simply the difference between the initial state in $^{22}Mg$ and the final state in $^{20}Ne$. Thus the branching ratios of the transitions from the same initial state to the same final state were summed and tabulated in table~\ref{tab:cal_br_2p}. For both cases, the two transitions which can be detected are emitted from the calculated IAS in $^{22}Mg$ to the $^{20}Ne$ ground state and to the $^{20}Ne$ first excited state. In the $4^+$ case, the transition to the ground state is weaker than the transition to the first excited state while the opposite is calculated for the $3^+$ case.   

\begin{table}
\begin{center}
\begin{tabular}{c|c|c|c|c}
     & Initial & Final & Transition  & Br(\%)\\
\noalign{\smallskip}\hline\noalign{\smallskip}   				 $4^+$& 13910		  & 0  	      	     & 5980  & 0.26  	   \\
     & 13910		  & 1765	     & 4215  & 1.19  	   \\
     & 13910		  & 4206	     & 1774  & 1.30 $10^{-2}$  \\
\hline                                     
$3^+$&  14101		  & 0		     & 6171  & 0.61	      \\
     &  14101  	  	  & 1765	     & 4406  & 0.36		\\
\hline                                     
\end{tabular}
\end{center}
\caption{\it Branching ratios for sequential $\beta$-2p emissions calculated for a ground state spin-parity of $4^+$ and $3^+$. 
For each initial state in $^{22}Mg$ and final state in $^{20}Ne$ the branching ratios of different transitions were summed. The energies tabulated are the calculated excitation energies. The total proton energy of the transition in the fourth column can be obtained by the subtraction of the final state energy, added to the 2p separation energy, from the initial state energy. All energies are in keV.}   
\label{tab:cal_br_2p}
\end{table}

\section{Results and discussion }
\label{sec:6}

\begin{figure*}
\begin{center}
\resizebox{0.75\textwidth}{!}{%
  \includegraphics{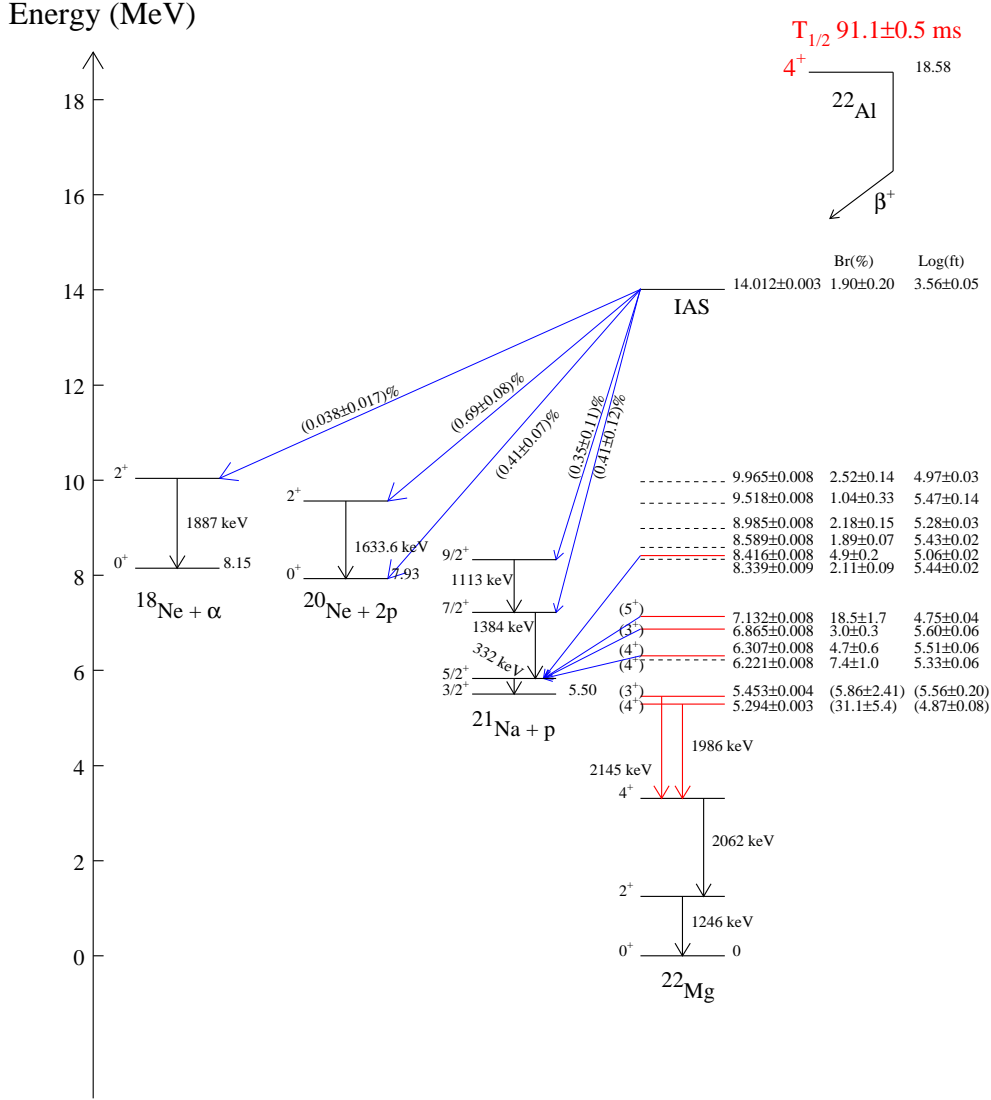}
}
\caption{\it $^{22}Al$ decay scheme deduced from the present experiment. Arrows are drawn for transitions with clearly identified coincidences with $\gamma$-rays except for the transition to the $^{20}Ne$ ground state. Levels supposed to emit protons to the $^{21}Na$ ground state are just drawn as dashed lines. 
Since no $\gamma$ coincidences have been determined for $\beta$-p transitions with a branching ratio less than 1\%, these transitions have not been drawn for the clarity of the scheme. All the energies, the branching ratios and the log(ft) of the excited states in $^{22}Mg$ determined from table~\ref{tab:pro22al} are tabulated in table~\ref{tab:Eex_22mg}. The branching ratios between brackets correspond to $\beta-\gamma$ decays deduced from the measured intensities of the $\gamma$-rays assuming the relative $\gamma$ intensities adopted (see reference~\cite{firestone05}).}
\label{fig:schema}
\end{center}
\end{figure*}

In this section, we compare our experimental results to previous measurements as well as to model predictions. This comparison 
will allow us to favour one of the two possible spin-parity assignments for the ground state of $^{22}Al$ and to construct a 
new $\beta$-decay scheme. \\

\begin{enumerate}

\item Half-life \\

The $^{22}Al$ half-life has been measured twice before this experiment. The first measurement was done by Cable {\it et al.}~\cite{cable82} and they found $T_{1/2}=70^{+50}_{-35}~ms$. The second measurement was performed by Blank {\it et al.}~\cite{blank}. 
The value measured was smaller $T_{1/2}=59 \pm 3~ms$ but consistent with the first value. This last value is in conflict with 
our value of $T_{1/2}=91.1 \pm 0.5~ms$. In reference~\cite{blank}, the half-life was measured in a beam-on/beam-off mode where 
the beam was switched off in a time interval of only 100~ms. If we assume a half-life close to 90~ms, this time interval is too 
short to correctly fit the decay curve when contamination is present. A fit was performed with the present data on a range of $100~ms$. 
The half-life obtained was $73.4 \pm 2.6~ms$. This indicates that performing a fit in such a short time interval reduced artificially 
the half-life value obtained, a fact also confirmed by a simulation.   

The values calculated with the $4^+$ (85~ms) and $3^+$ (80~ms) hypotheses are close to the present experimental value, slightly favouring 
a $4^+$ ground-state spin. However, no conclusion can be drawn from the half-life alone. \\

\item Experimental decay scheme \\

The informations extracted from the charged-particle spectrum (position of the peaks and their absolute intensity), from the measurement 
of $\gamma$ rays and the coincidences allow us to establish a new experimental $\beta$-decay scheme for $^{22}Al$ as shown in figure~\ref{fig:schema}. For transitions other than from the IAS, only those with a branching ratio higher than 1\% are shown. Otherwise, all the $^{22}Mg$ levels deduced from this experiment are tabulated in table~\ref{tab:Eex_22mg} with their branching ratios and log(ft) values. 

Starting from the energy measured in the E4 detector, the position of the levels in $^{22}Mg$ is calculated assuming that the levels are placed 
above the particle threshold, by adding the particle energy and the energy of the following $\gamma$ ray when observed. Thus, the transitions 1, 4, 6, 11 in table~\ref{tab:pro22al} in coincidence with the $\gamma$-ray at $332$ keV are coming from the levels situated at $6.307$, $6.865$, $7.132$ and $8.416$ MeV, respectively. 
The transition 20 is in coincidence with the $\gamma$-ray at $1384$ keV but no coincidences were clearly established with the $\gamma$-rays at $1113$ keV and $332$ keV. However, the energy level determined by adding the transition energy above the proton threshold to the energy of these three $\gamma$-rays leads to an energy compatible with the energy of the IAS. From this energy consideration, we assign the line 20 to a transition from the IAS to the $^{21}Na$ third excited state.

For the transition 24 at $6.774$ MeV, no $\gamma$-rays coincidences were determined but if we add the energy of the $\gamma$-rays $332$ keV and $1384$ keV, we obtain an energy of a $^{22}Mg$ level in agreement with the energy of the IAS. In addition, the shell model calculations predict a proton transition of $6.637$ MeV from the IAS to the $^{21}Na$ second excited state. So the line 24 can be assigned to a transition from the IAS to the $^{21}Na$ second excited state. 

Finally, as shown in figure~\ref{fig:schema}, five transitions can be attributed to an emission from the IAS in $^{22}Mg$. The IAS energy is determined as the mean value of level energies deduced from these five measured transitions. The $\beta$ feeding to the IAS is determined as the sum of the branching ratios of these transitions. \\

\begin{table}[htb]
\begin{center}
\begin{tabular}{|c|c|c|}
\hline
 $^{22}Mg$ levels ($MeV$)& $Br$ ($\%$)  &  Log(ft)\\
\hline
 5.294 $\pm$ 0.003 & 31.1 $\pm$ 5.4   &  4.87 $\pm$ 0.08 \\
\hline
 5.453 $\pm$ 0.004 & 5.86 $\pm$ 2.41  &  5.56 $\pm$ 0.20\\
\hline
 6.221 $\pm$ 0.008 & 7.39 $\pm$ 1.01    &  5.33 $\pm$ 0.06 \\
\hline
 6.307 $\pm$ 0.008 & 4.73 $\pm$ 0.63    &  5.51$\pm$ 0.06 \\
\hline
 6.476 $\pm$ 0.008 & 0.25$\pm$ 0.05   &  6.75 $\pm$ 0.09  \\
\hline 
 6.724 $\pm$0.008  & 0.75 $\pm$ 0.10  &  6.23 $\pm$ 0.06 \\
\hline
 6.865 $\pm$0.008  & 3.00 $\pm$ 0.34    &  5.60 $\pm$ 0.05 \\
\hline
 7.052 $\pm$0.008  & 0.81 $\pm$ 0.16  &  6.13 $\pm$ 0.09     \\
\hline
 7.132 $\pm$0.008  & 18.51 $\pm$ 1.74 &  4.75 $\pm$ 0.04 \\
\hline
 7.254 $\pm$0.008  & 0.45$\pm$ 0.08   &  6.34 $\pm$ 0.08             \\
\hline
 7.573 $\pm$0.008  & 0.48$\pm$ 0.07   &  6.25 $\pm$ 0.07      \\
\hline
 8.004 $\pm$0.009  & 0.64 $\pm$0.13   &  6.03  $\pm$ 0.09 \\
\hline
 8.339 $\pm$0.009  & 2.11 $\pm$0.09  &  5.44 $\pm$ 0.02	 \\
\hline
 8.416 $\pm$0.008 & 4.89 $\pm$ 0.24  &  5.06 $\pm$ 0.02 	 \\
\hline
 8.589 $\pm$0.008 & 1.89$\pm$0.07    &  5.43 $\pm$ 0.02	 \\
\hline
 8.985 $\pm$0.008 & 2.18$\pm$0.15    &  5.28 $\pm$ 0.03  	 \\
\hline 
 9.518 $\pm$0.008 & 1.04 $\pm$ 0.33  & 5.47$\pm$ 0.14		 \\
\hline
 9.725 $\pm$0.010 & 0.84 $\pm$ 0.11 &  5.51 $\pm$ 0.06            \\
\hline
 9.965 $\pm$0.008 & 2.52$\pm$0.14   &  4.97 $\pm$ 0.03 \\
\hline
 10.413 $\pm$0.010& 0.28$\pm$0.32 & 5.80 $\pm$ 0.50    \\
\hline
 10.678 $\pm$0.012& 0.29$\pm$0.11  & 5.70 $\pm$ 0.17 \\
\hline 
 11.309 $\pm$0.049& 0.18$\pm$0.55  & 5.70 $\pm$ 1.40   \\
\hline 
 11.410 $\pm$0.008& 0.21$\pm$0.62   & 5.60 $\pm$ 1.30    \\
\hline
 13.018 $\pm$0.056  & 0.33$\pm$0.07 &  4.81 $\pm$ 0.10	\\
\hline
 IAS: 14.012 $\pm$ 0.003 & 1.90 $\pm$0.20  & 3.56 $\pm$ 0.05   \\
\hline
\end{tabular}
\end{center}
\caption{\it Energy of the $^{22}Mg$ levels deduced from the $^{22}Al$ $\beta$ decay. The two first levels were determined from $\beta-\gamma$ transitions. The other ones were determined from the delayed charged particles detected and the $\gamma$-rays in coincidence when observed. For each level, the branching ratio from $^{22}Al$ is given as well as the log(ft) values calculated using the program available at~\cite{nndc}. The IAS energy was calculated as the mean value of all the transitions coming from this level.  }     
\label{tab:Eex_22mg}
\end{table}

\item Particle emission \\

In figure~\ref{fig:sum_br}, the summed branching ratios for charged-particle emission is shown as a function of the proton energy transition. This sum allowed us to compare the experimental results with the calculations without any interpretation of the data. The comparison concerns only the branching ratio related to a transition without any assignment to a level. This is not possible in the case of the summed Gamow-Teller strength where a level scheme have to be assumed before comparison.

The full points represent the experimental sum, while the lines represent the calculations performed with a OB interaction for both $3^+$ and $4^+$ spin-parities. The theoretical sum is limited to branching ratios larger than 0.25\% which is our experimental limit. There is a significant difference between the $3^+$ and $4^+$ case. The calculation for the $4^+$ case is in much better agreement with the experiment. \\ 

For the $\beta$-2p emission, two transitions were established in this experiment: to the $^{20}Ne$ ground state, a transition of 6085~keV with a branching ratio of $0.41 \pm 0.07~\%$ and to the $^{20}Ne$ first excited state, a transition of 4464 keV with a branching ratio of $0.69 \pm 0.08~\%$. According to table~\ref{tab:cal_br_2p} the calculated transition energies for the $3^+$ case are closer to experiment, but in the $4^+$ case the branching ratio to the $^{20}Ne$ ground state is weaker than the branching to the $^{20}Ne$ first excited state as measured in the experiment. The calculated branching ratios in both cases are of the same order of magnitude. The comparison of the $\beta$-2p emission with the calculation do not give additional information on the spin-parity of $^{22}Al$ ground state. However, since the transition energies and the branching ratios calculated are close to the measured ones, we can conclude that the $\beta$-2p emission observed is probably mainly a sequential emission. \\

\begin{figure}
\resizebox{0.5\textwidth}{!}{%
  \includegraphics{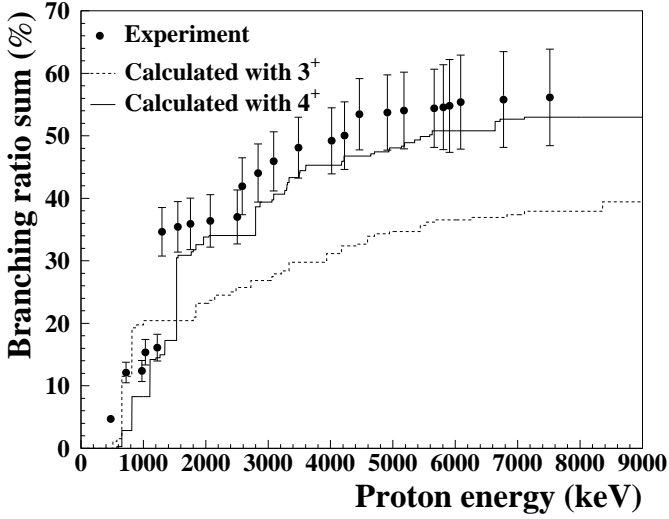}
}
\caption{\it Summed delayed proton branching ratios. The full circles represent the experimental values, the full line shows the calculated values with a ground-state spin-parity of $4^+$ and the dashed line is the result calculated with a ground-state spin-parity of $3^+$. Both calculations were performed with the OB interaction. Only branching ratios above 0.25\% were considered for the calculated sums.}
\label{fig:sum_br}
\end{figure}

\begin{figure}
\resizebox{0.5\textwidth}{!}{%
\includegraphics{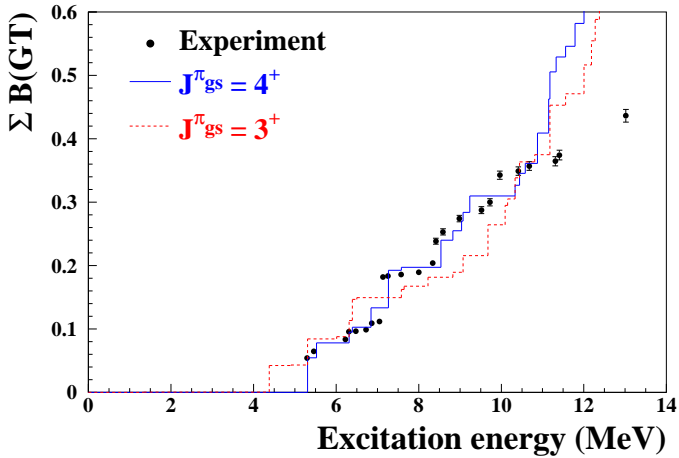}}
\caption{\it Summed Gamow-Teller strength for the $\beta$ decay of $^{22}Al$. The black circles represent the experimental values. 
The full line shows the shell-model results with a ground-state spin-parity of $4^+$, whereas the dashed line is due to a calculation with a ground-state spin-parity of $3^+$. Both calculations were performed with the OB interaction. }
\label{fig:BGT}
\end{figure}

\item Gamow-Teller strength \\

Figure \ref{fig:BGT} shows that the calculated summed Gamow-Teller strength in the case of a $4^+$ ground state is in better agreement with the experimental values than in the case of a $3^+$ ground state. The experimental strength was deduced from the experimental decay scheme established in the present work. The theoretical sum is obtained from branching ratios above 0.25\%. 
For higher energies, the experimental values are lower than the calculated sum. 
This is probably due to an incomplete measurement of $\beta$ decays to higher-lying states where the density of states increases and the branching ratios are small. Since figure~\ref{fig:sum_br} does not exhibit such kind of difference, it is also possible that an error in the interpretation of the origin of the experimental decay peaks contributes to this disagreement at higher energies.   

We do not observe a significant difference between the Gamow-Teller strength calculated using the USD or the OB interaction. The comparison of the experimental and the theoretical Gamow-Teller strength shows that the effective matrix elements developed by Brown and Wildenthal~\cite{brown85} for the mass region A=17-39 reproduce well the Gamow-Teller strength of $^{22}Al$. \\

\item The ground-state spin-parity of $^{22}Al$  \\

Prior to the present work, no convincing determination of the $^{22}Al$ ground-state spin was presented. 
The mirror nucleus $^{22}F$ has a $4^+$ ground state, but also a low-lying $3^+$ state. The shell model 
calculations suggest two main decays from $^{22}Al$ to $^{22}Mg$ levels below the proton threshold. In 
the case of a $3^+$ ground state spin-parity, $^{22}Al$ decays to a $^{22}Mg$ $2^+$ level at $4.401$ MeV 
and to a $4^+$ level at $5.311$ MeV. In the case of a $4^+$ ground state spin-parity, it decays to a 
$^{22}Mg$ $4^+$ level at $5.311$ MeV, to a $3^+$ level at $5.525$ MeV and weakly to a $4^+$ level at 
$3.313$ MeV as shown by the decay scheme in the middle of figure~\ref{fig:bet_gam_4} (the calculation 
using a $3^+$ value for the $^{22}Al$ ground state is not shown on this figure). The mirror nucleus $^{22}F$ 
decays to two $^{22}Ne$ $4^+$ levels and to a $^{22}Ne$ $3^+$ level in the same range of excitation energy 
as shown by the partial $\beta-\gamma$ decay scheme on the left of figure~\ref{fig:bet_gam_4}. 

Experimental results show two $\gamma$-rays at $1985.6$ keV and $2145$ keV. Two levels can be deduced from these measurements at $5.294$ MeV and $5.453$ MeV as shown on the right scheme of figure~\ref{fig:bet_gam_4}.  

 A simple comparison between these decay schemes fav-ours again the $4^+$ case because both $^{22}Al$ and $^{22}F$ do not decay to a $2^+$ level around $4$ MeV as calculated for the $3^+$ case.  

\begin{figure*}
\begin{center}
\resizebox{0.75\textwidth}{!}{%
  \includegraphics{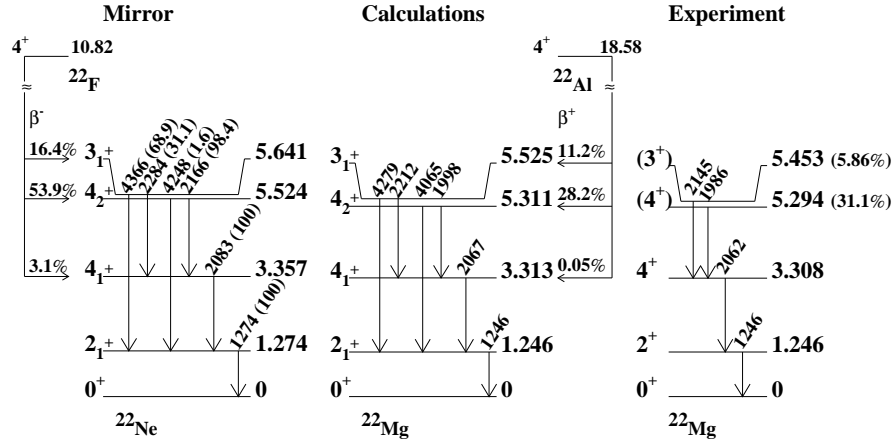}
}
\caption{\it Comparison of the partial $\beta$-$\gamma$ decay scheme of $^{22}F$ -the mirror nucleus of $^{22}Al$- (on the left) with the calculated decay scheme obtained with a spin-parity of $4^+$ for $^{22}Al$ ground state (in the middle) and the decay scheme deduced from this experiment (on the right). The partial $\beta$-$\gamma$ decay scheme of $^{22}F$ was taken from the reference~\cite{firestone05}. }
\label{fig:bet_gam_4}
\end{center}
\end{figure*}

Another way to analyse these results is to compare quantitatively the absolute $\gamma$-ray intensities measured and the intensities deduced from shell model calculation. The $\gamma$-ray intensities are calculated as the product of the calculated $\beta$-decay branching ratios from the $^{22}Al$ ground state to the $^{22}Mg$ levels and the relative intensities of the $\gamma$ 
ray from these levels. The relative intensities are assumed to be the same as the intensities of the analogue transitions in the $^{22}Ne$ mirror nucleus.

In figure~\ref{fig:intens_gam}, the calculated intensities of the expected $\gamma$ rays - taking into account 
our limited range of $\gamma$ energies - for both cases are compared to the experimental results. We notice that 
only three $\gamma$-rays are predicted in the $3^+$ case whereas four $\gamma$-rays are predicted in the $4^+$ 
case as for the experimental results. A much better agreement is observed between the calculated intensities for 
a $4^+$ spin-parity and our experimental results. \\

Therefore, from the different pieces of evidence previously discussed, we deduce that the $^{22}Al$ ground state 
spin-parity is most likely $4^+$. \\

\item $^{22}Mg$ levels assignment \\

Here we assume a 4$^{+}$ spin assignment for the ground state of
$^{22}$Al. With this condition, the following levels in $^{22}$Mg
are discussed and are tentatively assigned:
\begin{itemize}
\item E$_{x}$ = 5.294 MeV. Experimentally we concluded that the
observed $\gamma$ ray at $1985.6$ keV is compatible with the adopted $\gamma$ ray at $1984.8$ keV which is emitted from a level situated at $5.293$ MeV with a spin-parity from $3^+$ to $5^+$~\cite{firestone05}. In addition, the comparison between the $^{22}Mg$ and $^{22}Ne$ $\gamma$ decay schemes and the shell model calculations (see figure~\ref{fig:bet_gam_4}) shows that this ray is probably emitted from a $4^+$ level. So we can establish that the level measured at $5.294 \pm 0.003$ MeV is a $4^+$ one.

\item E$_{x}$ = $5.453$ MeV. There is a known state positioned at
E$_{x}$ = 5.452 MeV with a spin-prity $2^+$ or $3^+$~\cite{firestone05}. A comparison with the mirror decay scheme and the shell model calculations (figure~\ref{fig:bet_gam_4}) allowed us to assign a spin $3^+$ for this level.

\item E$_{x}$ = $6.221$ MeV. This level is fed with a branching
ratio of 7.4\% and a log(ft) of 5.32. Shell model calculations predict the existence of a proton transition from a $4^+$ level at $6.315$ MeV with a
branching ratio of 5.74\%. In the mirror nucleus $^{22}Ne$, a $4^+$ level is situated at $6.346$ MeV and is fed by $\beta^-$ decay with a branching ratio of 7.0\% and a log(ft) of 5.34. These facts are strong evidences that this level has spin $4^+$.

\item E$_{x}$ = $6.307$ MeV and E$_{x}$ = $6.865$ MeV. These
levels have branching ratios of 4.7\% and 3.0\%. Shell model
calculations predict the existence of only two levels in this
range of energy, positioned at $6.394$ MeV and $6.844$ MeV, with
spin $3^+$ and $4^+$ and with branching ratios 2.67\% and 8.24\%.
There are evidences (see \cite{smirnova}) that
the first state corresponds to a $4^+$ state measured at
E$_{x}$ = $6.267$ MeV. In this case, the second state is assigned
to a spin-parity $3^+$.

\item E$_{x}$ = $7.132$ MeV. The strongest proton transition is
predicted to originate from a $5^+$ level at $7.270$ MeV with a
branching ratio of 13.2\% while experimentally it decays from a
level at $7.132$ MeV with a branching ratio of 18.51\% and a log(ft) of 4.75. 
In the mirror nucleus, a $5^+$ level is situated at $7.422$ MeV and is fed by $\beta^-$ decay with a branching ratio of 8.7\% but with a log(ft) of 4.7. It is the strongest $\beta^-$ feeding after the two levels corresponding to those decaying by $\gamma$ emission. So the level measured at $7.132$ MeV can be assigned to a spin-parity $5^+$. \\

\end{itemize}


\begin{figure}
\resizebox{0.5\textwidth}{!}{%
  \includegraphics{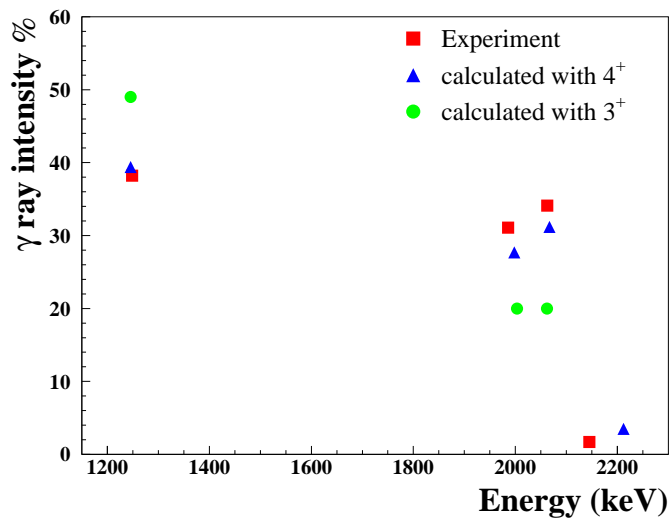}
}
\caption{\it Experimental $\beta$-$\gamma$ ray intensities (squares) are compared to the calculated ones in the case of a $4^+$ (triangles) and a $3^+$ (full circles) $^{22}Al$ ground-state spin-parity.}
\label{fig:intens_gam}
\end{figure}
 
\item Mass excess of $^{22}Al$ \\

 The mass excess of the $^{22}Al$ ground state can be deduced from the measurement of the $^{22}Mg$ IAS energy. Indeed, four 
states are experimentally observed for this T=2 isobaric multiplet: 
 
\begin{enumerate}

\item the $^{22}Al$ ground state with $T_Z=-2$.

\item the $^{22}Mg$ excited state we measured at an energy of $14011.7 \pm 3.5$~keV with $T_Z=-1$ and a mass excess of $-396.8 \pm 1.4$~keV~\cite{audi97}. 

\item the $^{22}Ne$ excited state measured at an energy of $14070 \pm 40$~keV with $T_Z=+1$ and a mass excess of $-8024.3 \pm 0.22$~keV~\cite{audi97}. 

\item the $^{22}F$ ground state with $T_Z=+2$ and a mass excess of $2794 \pm 12$~keV~\cite{audi97}.

\end{enumerate}

The last three states are experimentally measured and permit the determination of the IMME parameters. This equation gives the mass 
of states which belong to the same isospin multiplet: $M(A,T,T_Z) = a(T,Z) + b(A,T)T_Z + c(A,T)T_Z^2$. By using the mass excess of the 
three known states, we calculated the IMME coefficients as follows: $a= 9652.7 \pm 40.2$, $b=-3784.6 \pm 20.1$ and $c=177.6 \pm 20.4$. 
With these values, we determined the mass of the $T_Z=-2$ nucleus $^{22}Al$ to be $17932 \pm 99$ keV. This mass excess is close to the 
extrapolation from Audi {\it et al.}~\cite{audi03} which predicts 
a mass excess of $18180 \pm 90$ keV. \\

\item Mirror asymmetry \\

The spin-parity assignment of four levels in $^{22}Mg$ allowed us to determine the asymmetry factor $\delta$ for analogue 
$\beta$ decays to $^{22}Mg$ and to $^{22}Ne$ states. This parameter is calculated as: 
\begin{center} 
$\delta =(ft^+/ft^-) - 1$
\end{center}
f is the Fermi function and t the partial half-life, +(-) stands for $\beta^{+(-)}$ decays.
Table \ref{tab:asym} gives the asymmetry parameters determined for the decay to $^{22}Mg$ and $^{22}Ne$ mirror levels. 
The error bars of these numbers are relatively large especially for the $3^+$ state due to rather large uncertainty for the $\gamma$-ray intensity. No conclusive information can be drawn from these values. Nevertheless, for the other states, the asymmetry factors are in agreement with the values determined in the $sd$ shell.

\begin{table}
\begin{center}
\begin{tabular}{|c|c|c|r|}
\hline
spin   & $^{22}$Mg       & $^{22}$Ne     & \multicolumn{1}{c|}{$\delta$}         \\
\hline
 $4^+$ &   5.294~MeV    &  5.524~MeV	 & 0.202$\pm$0.037    \\
 $3^+$ &   5.453~MeV    &  5.641~MeV	 & 0.995$\pm$0.460    \\
 $4^+$ &   6.221~MeV    &  6.346~MeV	 & -0.023$\pm$0.003    \\
 $5^+$ &   7.132~MeV    &  7.422~MeV	 & 0.122$\pm$0.014    \\
\hline
\end{tabular}
\caption{\it Asymmetry parameters $\delta$ of several analogue levels in $^{22}$Mg and $^{22}$Ne fed in the $\beta$ decays of $^{22}$Al and $^{22}$F mirror nuclei.}     
\label{tab:asym}
\end{center}
\end{table}


\end{enumerate}

\section{Conclusion}
\label{sec:8}

In the present work, the decay of $^{22}Al$ was studied after its implantation in a silicon detector. 
The nuclei were produced by fragmentation of a $^{36}Ar$ primary beam in a carbon target and selected by the LISE3 facility at GANIL. 
The high purity of the secondary beam (93\%) permits a precise measurement of the $^{22}Al$ half-life (91.1$\pm$0.5 ms) and a good energy resolution for the measurement of $\beta$-delayed particle transitions. 

The measurement of $\gamma$ rays has confirmed a $\beta-2p$ decay branch to the $^{20}Ne$ first excited state and a $\beta$-$\alpha$ 
decay to the $^{18}Ne$ first excited state. In addition, it allowed for the spin-parity assignment of six levels in $^{22}Mg$.

A complete calculation of the $\beta$-delayed particle decay was performed using shell model calculations in the $sd$ shell. The comparison 
of the theoretical summed Gamow-Teller strength and the experimental values shows good agreement. As a conclusion, the $^{22}Al$ $\beta$ decay 
is well described by the matrix elements developed by Brown and Wildenthal~\cite{brown85} where the quenching factor is included. 

The $^{22}Al$ ground-state spin-parity was assigned to be most likely $4^+$. This assignment is based on a comparison of the present experimental results with theoretical calculations and the mirror nucleus. The measurement of the $\gamma$ rays and the Gamow-Teller strength distribution was essential to reach this conclusion.

\section*{Acknowledgment}

We would like to thank the GANIL and particularly the LISE technical staff for their help during this experiment and our colleagues from the EXOGAM collaboration for kindly providing us with the clover detector used here.
This work was supported in part by the Conseil r\'egional d'Aquitaine.

\bibliography{biblio_22al}

\end{document}